\documentclass[pra,twocolumn,amsmath,amssymb,floatfix,reprint,footinbib,superscriptaddress,longbibliography,showkeys]{revtex4-2}
\usepackage{picinpar,graphicx}
\usepackage{braket}
\usepackage{amsmath}
\usepackage{graphicx}
\usepackage{epstopdf}
\usepackage{dcolumn}
\usepackage{graphicx}
\usepackage{mathrsfs}
\usepackage{mdwlist}

\usepackage{placeins}
\usepackage{subfigure}
\usepackage{booktabs}
\usepackage{amsmath}
\usepackage{dsfont}
\usepackage{amstext}
\usepackage{amssymb}
\usepackage{amsbsy}
\usepackage{bbm}
\usepackage{amsthm}
\usepackage{graphicx}
\usepackage{textcomp}
\usepackage{multirow}
\usepackage{color}
\usepackage[colorlinks,citecolor=blue]{hyperref}
\definecolor{Dgreen}{RGB}{0, 100, 0}
\usepackage{url}
\usepackage[colorlinks]{hyperref}
\hypersetup{%
	plainpages=true,
	breaklinks=true,  
	hypertexnames=false,  
	pageanchor=true,
	colorlinks=true,
	linkcolor={blue},
	citecolor={red},
	urlcolor={blue},
	anchorcolor={black}
}

\usepackage{xcolor}
\usepackage[normalem]{ulem} 
\begin{document}

	\title{Reinforcement Learning for Fast and Robust Longitudinal Qubit Readout}
	\author{Yiming Yu}
    \affiliation{Fujian Key Laboratory of Quantum Information and Quantum Optics, College of Physics and Information Engineering, Fuzhou University, Fuzhou 350108, China}

	\author{Yuan Qiu}
    \affiliation{Fujian Key Laboratory of Quantum Information and Quantum Optics, College of Physics and Information Engineering, Fuzhou University, Fuzhou 350108, China}

    \author{Xinyu Zhao}
    \affiliation{Fujian Key Laboratory of Quantum Information and Quantum Optics, College of Physics and Information Engineering, Fuzhou University, Fuzhou 350108, China}
	
	\author{Ye-Hong Chen}\thanks{yehong.chen@fzu.edu.cn}
    \affiliation{Fujian Key Laboratory of Quantum Information and Quantum Optics, College of Physics and Information Engineering, Fuzhou University, Fuzhou 350108, China}
	\affiliation{Quantum Information Physics Theory Research Team, Center for Quantum Computing, RIKEN, Wako-shi, Saitama 351-0198, Japan}
	
	\author{Yan Xia}\thanks{xia-208@163.com}
    \affiliation{Fujian Key Laboratory of Quantum Information and Quantum Optics, College of Physics and Information Engineering, Fuzhou University, Fuzhou 350108, China}

	\date{\today}
	
	\begin{abstract}
		Longitudinal coupling offers a compelling pathway for quantum nondemolition (QND) readout, but pulse design is constrained by hardware limitations such as the coupling strength and the photon number required to stay within the linear regime. 
        We develop a reinforcement  learning framework to optimize the longitudinal coupling waveform under such constraints. 
        Building upon the theoretical foundation of shortcuts to adiabaticity (STA), we parameterize an auxiliary trajectory with cubic B-splines and reconstruct the physical control. 
        At a fixed short readout time, the optimized pulse converges to a constraint saturating flat-top protocol and yields a approximately $50\%$ improvement in $\mathrm{SNR}$ over an STA baseline, while exhibiting enhanced robustness to parameter drifts. 
        Simulation results demonstrate the efficacy of reinforcement learning in optimizing longitudinal readout pulses. 
        The optimized protocol attains substantial performance gains and yields smooth, hardware-compatible waveforms governed by an interpretable ``saturate-and-hold'' mechanism.
	\end{abstract}

	\maketitle

\section{INTRODUCTION}
Fast, high-fidelity, single-shot readout is a prerequisite for scalable quantum information processing, particularly for implementing quantum error correction and measurement-based feedback~\cite{PhysRevLett.93.207002,RevModPhys.82.1155}.
In quantum electrodynamics (QED), dispersive readout provides the canonical route to quantum nondemolition (QND) measurement~\cite{Blais2004cQED, Wallraff2004StrongCoupling,Krantz2019QuantumEngineersGuide, Blais2021cQEDReview,Schuster2007ResolvingPhotonNumber,PhysRevX.10.011045,Spring2025PRXQFastMultiplexedPurcell}. 
By operating in the large-detuning regime, the qubit imparts a state-dependent shift to the resonator frequency; this response is subsequently extracted via homodyne detection of the cavity field.
However, dispersive readout is constrained by a trade-off between faster signal-to-noise
ratio (SNR) acquisition and the suppression of measurement-induced qubit transitions~\cite{PhysRevA.76.012325,PhysRevLett.106.110502,PhysRevLett.109.240502, PhysRevLett.133.233605,PhysRevA.74.042318,PhysRevApplied.7.054020}. 
In particular, increasing the drive strength to improve the SNR often excites the system beyond the critical photon number, leading to non-adiabatic transitions that compromise the QND nature of the measurement\cite{5015954,npjQI2023ThresholdNoQLA}.

To overcome the above limitation, longitudinal readout has been proposed as a promising alternative~\cite{PhysRevLett.115.203601,PhysRevB.93.134501}. 
Specifically, longitudinal readout relies on a $\sigma_z$ qubit–resonator interaction. 
Its coupling strength acts as a state dependent drive that displaces the cavity field, thereby encoding the qubit state into the amplitude of the output field~\cite{PhysRevLett.115.203601,PhysRevApplied.23.034067}. 
In the ideal limit, this interaction constitutes a QND measurement because it commutes with the qubit Hamiltonian, enabling faster information extraction with relaxed bandwidth constraints compared to dispersive readout~\cite{Lupascu2007QND,Krantz2019QuantumEngineersGuide}.
Recent advances~\cite{Ye2024UltrafastQuartonReadout,PhysRevLett.134.037003,98n9-13y4,PhysRevApplied.18.034010,PhysRevLett.122.080502,PhysRevApplied.17.064006,mp2x-zj3y} have solidified this paradigm.
In particular, the quarton coupler demonstrates ultrafast, purcell-free readout enabled by potential engineering~\cite{Ye2024UltrafastQuartonReadout}.
A unified Floquet framework reveals the intrinsic link between longitudinal and dispersive mechanisms through the driven qubit's AC Stark response, and clarifies how longitudinal modulation can offer faster, more QND-like signal extraction~\cite{PhysRevLett.134.037003}.

Realizing the full potential of longitudinal readout requires precise pulse engineering~\cite{PhysRevLett.122.080502,PhysRevLett.122.080503,Ran:20}. 
Numerical optimal control methods, such as GRAPE~\cite{Khaneja2005GRAPE} and Krotov~\cite{Reich2012Krotov} method, offer the flexibility to explore broad parameter spaces. 
However, enforcing multiple simultaneous rigid constraints remains challenging for gradient-based optimal control. 
In particular, strict saturation bounds on the coupling strength impose rigid amplitude limits on the control field~\cite{PhysRevLett.103.110501, PhysRevApplied.5.011001}. 
In addition, upper bounds on the instantaneous intracavity photon number make the optimization landscape highly nonconvex, which often leads to poor convergence or experimentally infeasible pulses.

Reinforcement learning (RL) has recently emerged as a powerful tool for quantum control, capable of optimizing nonconvex objectives under rigid constraints~\cite{Bukov2018RL, Fosel2018RL,Li2025RLfDnpjQI,PhysRevApplied.21.044012,PhysRevA.108.032616,7rc4-p446}. 
However, in pulse-shaping problems the action space is extremely complex~\cite{niu2019universal, PhysRevX.12.011059}.
Without a structured exploration space, the agent often expends many updates on unphysical or hardware-infeasible waveforms, leading to slow convergence and unstable training.
In practice, RL needs a physics-based starting point (and a constrained parameterization) so that exploration stays inside the feasible region from the beginning.

In this work, we employ RL to optimize longitudinal readout pulses based on a single mode effective model. Building on shortcuts to adiabaticity (STA) and inverse engineering~\cite{PhysRevApplied.18.034010,RevModPhys.91.045001,PhysRevLett.126.023602,PhysRevA.93.052109,PhysRevLett.104.063002,Yu:25,PhysRevA.94.052311,PhysRevA.101.023822,PhysRevLett.126.023602}, we start from an STA-derived analytical seed pulse~\cite{PhysRevLett.118.100601}.
This inverse-engineering parameterization guarantees smooth boundary conditions by construction~\cite{PhysRevA.82.053403}, enabling the agent to focus exclusively on shaping the pulse to maximize the SNR. 
Coupled with the physics-based initialization, this focused exploration drastically reduces the number of training iterations required to reach convergence.

Our optimization framework utilizes Proximal Policy Optimization (PPO)~\cite{schulman2017proximalpolicyoptimizationalgorithms} to search for feasible pulse shapes, integrating hardware constraints directly into the control loop.
Our simulation results compare the proposed approach with (i) an STA-only baseline~\cite{PhysRevApplied.18.034010} and (ii) an RL baseline without a physics-based initialization. 
The proposed method achieves higher final SNR and improved robustness to parameter drift, highlighting the benefit of combining RL with a physically grounded starting point. 
We further test all methods under several hardware constraint settings (e.g., different limits on the coupling amplitude and the intracavity photon number). 
Across these settings, our approach consistently achieves the best SNR while remaining within the allowed operating range.

\section{Effective model and control parameterization}\label{s2}

\subsection{Inverse Engineering Control Model for Longitudinal Readout}\label{s2.1}

\begin{figure}[t] 
    \centering
    \includegraphics[width=0.47\textwidth]{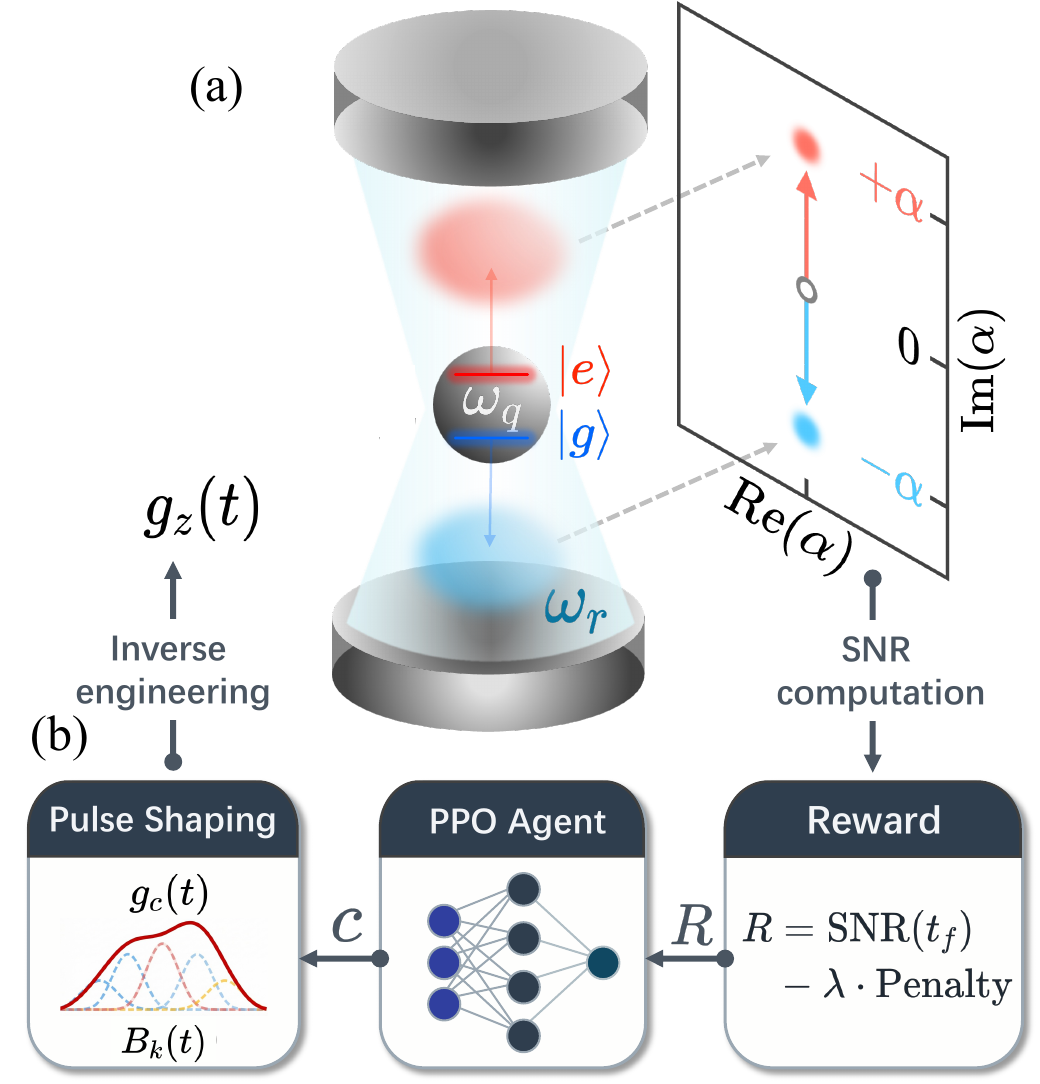} 
    \caption{Schematic of the physics-based reinforcement learning framework for optimizing longitudinal qubit readout.
    (a) Physical implementation of the longitudinal readout. A qubit (central sphere) is coupled to a resonator mode with frequency $\omega_r$ via a time-dependent longitudinal interaction $g_z(t)\sigma_z(\hat{a} + \hat{a}^\dagger)$. Depending on the qubit state $|e\rangle$ (red) or $|g\rangle$ (blue), the cavity field is displaced to symmetric conditional pointer states $+\alpha$ and $-\alpha$ in the phase space, spanned by the expectation value of the annihilation operator $\langle \hat{a} \rangle$.
    (b) The closed-loop optimization control flow. The PPO agent outputs a set of coefficients $\mathbf{c}=\{c_k\}$ to construct the auxiliary control trajectory $g_c(t)$ using cubic B-spline basis functions $B_k(x)$. The physical control pulse $g_z(t)$ is then derived via inverse engineering constraints to ensure smooth boundary conditions. The system performance is evaluated to generate a reward $R$, which maximizes the signal-to-noise ratio (SNR) at the readout time $t_f$ while penalizing physical constraint violations.}
    \label{fig:framework}
\end{figure}

A QED system consisting of a qubit coupled to a single-mode readout resonator is considered~\cite{RevModPhys.93.025005,doi:10.1126/science.1231930,PhysRevA.76.042319}. To achieve fast and high-fidelity readout, the longitudinal coupling mechanism is utilized, which employs the computational basis states of qubit to directly drive the displacement of the resonator field. Upon truncating the qubit to its lowest two energy levels ($\{|g\rangle, |e\rangle\}$), the system Hamiltonian is given by:
\begin{equation}
H(t) = \omega_r \hat{a}^\dagger \hat{a} + \frac{\omega_q}{2}\sigma_z + g_z(t)\,\sigma_z\,(\hat{a} + \hat{a}^\dagger),
\label{eq:hamiltonian}
\end{equation}
where $\hat{a}$ ($\hat{a}^\dagger$) is the annihilation (creation) operator of the resonator mode with frequency $\omega_r$, $\omega_q$ is the qubit transition frequency, $\sigma_z$ is the Pauli-$Z$ operator, and $g_z(t)$ represents the time-dependent physical longitudinal coupling strength.
The physical model illustrate in the Fig.~\ref{fig:framework}(a).
This effective Hamiltonian can be physically realized in superconducting circuit architectures by employing Josephson-junction-based couplers~\cite{PhysRevB.93.134501,PhysRevLett.115.203601}.

This Hamiltonian describes a state-dependent displacement of the oscillator. 
The qubit state determines the trajectory of the cavity field in phase space, creating the separation required for high-fidelity readout. 
Because the interaction term commutes with the free Hamiltonian, this process represents a QND measurement~\cite{PhysRevLett.115.203601}.

To overcome the limitations of adiabatic pulse shaping and speed up QND readout, we adopt an inverse engineering approach within the STA framework to design pulse~\cite{PhysRevApplied.18.034010}. 
By working in the resonator rotating frame and applying a time-dependent displacement transformation, the longitudinal coupling strength $g_z(t)$ can be expressed in terms of a smooth auxiliary trajectory $g_c(t)$:
\begin{equation}
    g_z(t) = g_c(t) + \frac{1}{\omega_r^2}\ddot{g}_c(t).
    \label{eq:inverse_engineering}
\end{equation}
Here, $g_c(t)$ has the same units as $g_z(t)$ and the dot denotes derivatives with respect to the physical time $t$. 
$g_c(t)$ defines the target displacement trajectory of the resonator field in phase space, thereby establishing a rigorous analytical relation to the physical control parameters~\cite{PhysRevApplied.18.034010}. 
The detailed derivation process are provided in Appendix~\ref{app:inverse_engineering}.

To ensure experimentally feasible waveforms with no initial jumps, $g_c(t)$ is required to satisfy the boundary conditions $g_c(0)=g_c(t_f)=0$ as well as vanishing first and second derivatives ($\dot{g}_c = \ddot{g}_c = 0$) at the boundaries~\cite{PhysRevA.97.013631}. 
Specifically, to counteract non-adiabatic transitions induced by fast driving, the physical coupling strength $g_z(t)$ must include a correction term to the second derivative of the $g_c(t)$.

\subsection{Signal Generation and Hardware-Constrained Optimization Landscape}\label{s2.2}

To rigorously quantify the signal generation process, we establish the link between the experimentally observable signals and the internal system dynamics using the standard input-output formalism~\cite{PhysRevLett.115.203601}. The output field operator $\hat a_{\mathrm{out}}(t)$, which determines the homodyne detection signal, relates to the intracavity field $\hat a(t)$ via the boundary condition:
$\hat a_{\mathrm{out}}(t) = \hat a_{\mathrm{in}}(t) + \sqrt{\kappa} \hat a(t),$
where $\hat a_{\mathrm{in}}$ denotes the input noise operator (assumed to be in the vacuum state) and $\kappa$ is the resonator energy decay rate. Consequently, the conditional expectation values of the output field map directly to the scaled intracavity amplitudes: \begin{equation}
\langle \hat a_{\mathrm{out}}(t) \rangle = \sqrt{\kappa} \langle \hat a(t) \rangle.
\label{eq:out}
\end{equation}

Working in the resonator rotating frame, the field is characterized by the conditional coherent amplitudes
\begin{equation}
    \alpha_j(t) \equiv \langle\hat{a}(t)\rangle_j,
\label{eq:alpha}
\end{equation}
where the index $j \in \{e, g\}$ denotes the qubit excited ($|e\rangle$) and ground ($|g\rangle$) states, respectively. Under the inverse engineering framework~\cite{RevModPhys.91.045001,PhysRevLett.105.123003,PhysRevA.97.023841}, these conditional amplitudes are uniquely determined by the auxiliary function $g_c(t)$:
\begin{equation}
\alpha_{e,g}(t) = \mp i\, e^{-\kappa t/2} \int_{0}^{t} g_c(\tau)\, e^{\kappa \tau/2}\, d\tau,
\label{eq:alpha_eg}
\end{equation}
where the minus (plus) sign corresponds to the excited (ground) state.

Due to the symmetry of the longitudinal drive, the trajectories satisfy $\alpha_{e}(t) = -\alpha_{g}(t)$. According to Eq. (\ref{eq:out}) and Eq. (\ref{eq:alpha}), the instantaneous pointer-state separation $d_{\mathrm{out}}(t)$ at the output is derived as:
\begin{equation}
d_{\mathrm{out}}(t) = \big| \langle \hat a_{\mathrm{out}}\rangle_e - \langle \hat a_{\mathrm{out}}\rangle_g \big| = 2\sqrt{\kappa}|\alpha_{e}(t)|,
\label{eq:dout}
\end{equation}
where $d_{\mathrm{out}}(t)$ directly determines the extractable signal strength. By combining Eq. (\ref{eq:alpha_eg}), the pulse optimization task is rephrased as a $g_c(t)$ trajectory planning problem aimed at maximizing $|\alpha_{j}(t)|$.

Readout performance is quantified using homodyne detection of the output field with overall measurement efficiency $\eta$. The homodyne measurement operator~\cite{RevModPhys.82.1155}, representing the weighted integration of the signal quadratures, is defined as:
\begin{equation}
\hat{\mathcal M}_\phi(t)
\equiv
\sqrt{\eta\kappa}\int_{0}^{t} d\tau\,
\Bigl(e^{-i\phi}\hat a_{\mathrm{out}}(\tau)+e^{i\phi}\hat a_{\mathrm{out}}^\dagger(\tau)\Bigr),
\label{eq:M_def}
\end{equation}
where \(\phi\) is the local-oscillator (LO) phase. The single-shot distinguishability between the two qubit states \(j \in \{e, g\}\) is characterized by the signal-to-noise ratio (SNR):
\begin{equation}
\mathrm{SNR}(t)=
\frac{\big|\langle \hat{\mathcal M}_\phi(t)\rangle_{e}-\langle \hat{\mathcal M}_\phi(t)\rangle_{g}\big|}
{\sqrt{\mathrm{Var}_{e}[\hat{\mathcal M}_\phi(t)]+\mathrm{Var}_{g}[\hat{\mathcal M}_\phi(t)]}},
\label{eq:snr_quantum_def}
\end{equation}
where \(\mathrm{Var}_{j}[\hat{\mathcal M}]\equiv\langle \hat{\mathcal M}^2\rangle_{j}-\langle \hat{\mathcal M}\rangle_{j}^2\) denotes the variance of the measurement record~\cite{RevModPhys.82.1155,PhysRevA.76.012325}.

Crucially, by optimizing the LO phase $\phi$ to align with the displacement direction, the signal difference in the numerator corresponds directly to the time-integrated pointer-state separation derived in Eq.~\eqref{eq:dout}:
\begin{equation}
\big|\langle \hat{\mathcal M}_\phi(t)\rangle_{e}-\langle \hat{\mathcal M}_\phi(t)\rangle_{g}\big| = \sqrt{\eta} \int_{0}^{t} d_{\mathrm{out}}(\tau)\, d\tau.
\label{eq:numerator_link}
\end{equation}
This relation explicitly connects the instantaneous separation $d_{\mathrm{out}}(t)$to the final accumulated fidelity. 
For numerical optimization, we evaluate the SNR using an equivalent compact expression derived under the linear coherent-state limit; the equivalence and conventions are provided in Appendix~\ref{app:snr_proof}.

The ultimate goal is to maximize the final SNR, $\mathrm{SNR}(t_f)$, within a fixed short readout time $t_f$. However, this optimization objective is bounded by strict physical constraints inherent to the inverse engineering model and hardware capabilities~\cite{PhysRevApplied.18.034010}. 
Specifically, Eq.~(\ref{eq:inverse_engineering}) reveals that the physical coupling $g_z(t)$ scales with the acceleration term $\ddot{g}_c(t)$.
Consequently, any rapid variations in the auxiliary trajectory $g_c(t)$ translate to large spikes in the required drive amplitude, risking instantaneous violations of the hardware saturation limit $g_{\max}$ or bandwidth constraints.
Furthermore, although the inverse engineering mapping assumes a linear resonator, excessive intracavity photon numbers $N(t)=|\alpha(t)|^2$ can induce parasitic self-Kerr nonlinearities.
Crucially, such nonlinearities would invalidate the linear analytical model derived in Sec.~\ref{s2.1}, rendering the calculated control pulses inaccurate and degrading readout fidelity.

\subsection{Reinforcement Learning Framework with Physics-Based Parameterization}\label{s2.3}

The above optimization problem evolves from simple integration maximization into a complex constrained trajectory planning problem.
An optimal path $g_c(t)$ must be identified within a highly non-convex parameter space defined by amplitude saturation, bandwidth limitations, and model-validity bounds. 
RL is particularly well-suited for this landscape, as it can flexibly encode these complex constraints into reward signals and autonomously navigate the dynamic boundaries to locate stable convergence to a high-performing solution.
To explore optimal control strategies beyond analytical solutions, the schematic workflow is depicted in the Fig.~\ref{fig:framework}(b).

Unlike standard RL tasks that require sequential decision-making at every time step~\cite{10.1561/2300000021}, our approach adopts a global parameterization strategy. 
This choice is fundamentally dictated by the physical nature of the inverse engineering framework in Eq.~(\ref{eq:inverse_engineering}), where the physical drive $g_z(t)$ is influenced by $\ddot g_c(t)$. 
A step-by-step generation of $g_c(t)$ would inject high-frequency components from stochastic network outputs; these components are amplified in $g_z(t)$ leading to experimentally infeasible oscillations.

In this framework, $g_c(t)$ is parameterized by cubic B-spline, yielding a $C^2$-smooth waveform with local shape control\cite{deBoor1978}:
\begin{equation}
g_c(t)=\sum_{k=1}^{N_{\mathrm{B}}} c_k\,B_k(t).
\end{equation}
Here $\{B_k\}$ are cubic B-spline basis functions, $N_{\mathrm{B}}$ is the number of basis functions, and $\mathbf{c}=\{c_k\}$ is the coefficient vector sampled by the policy in a single-step episode. The local support property of cubic B-splines ensures that adjusting a single control point only modifies a specific segment of the waveform, allowing for precise local shaping without unintentionally altering distant regions of the pulse.
To enforce these boundary conditions derived in Sec.~\ref{s2.1}, we fix the first and last three spline coefficients to zero, so that in our spline convention, $g_c$, $\dot g_c$, and $\ddot g_c$ vanish at $t=0$ and $t=t_f$.

While the physics simulation yields a deterministic outcome for any given set of coefficients $\mathbf{c}$, the RL optimization faces a significant challenge: the search space is filled with local optima and discontinuities due to the complex mapping and strict hardware constraints. 
In such a difficult regime, starting from scratch is inefficient. 
Instead, we initialize the policy using an STA solution. 
This anchors the agent in spline-coefficient space, constraining early exploration to valid physical trajectories and preventing the optimizer from wasting time in infeasible regions.

To optimize the policy parameters in this parameterized control setting, PPO is adopted. 
PPO’s clipped surrogate objective limits the policy change per update preventing destructive updates that would distort the seed waveform or incur large constraint violations, thereby promoting stable convergence to a high-SNR feasible solution.

The optimization is guided by a composite scalar reward function $R(\mathbf{c})$:

\begin{equation}
    \begin{aligned}
    &R(\mathbf{c})
    = R_{\mathrm{SNR}}(\mathbf{c}) - \sum_{j}\lambda_{j} P_{j},\\[2pt]
    &R_{\mathrm{SNR}}(\mathbf{c})
    = w_{\mathrm{tf}} \ln \tilde{S}(t_f)
     + w_{\mathrm{avg}} \ln \langle \tilde{S} \rangle_t,\\[2pt]
    &\tilde{S}(t)
    \equiv \mathrm{SNR}(t)+\epsilon,\qquad\langle \tilde{S} \rangle_t
    \equiv \frac{1}{t_f}\int_{0}^{t_f}\tilde{S}(t)\,\mathrm{d}t .
    \end{aligned}
    \end{equation}
$R_{\mathrm{SNR}}(\mathbf{c})$ promotes rapid information acquisition by combining the final SNR and its time-averaged value. $R_{\mathrm{SNR}}(\mathbf{c})$ combines the final-time performance $\tilde{S}(t_f)$ and its time average $\langle \tilde{S}\rangle_t$, weighted by $w_{\mathrm{tf}}$ and $w_{\mathrm{avg}}$, respectively, so that the optimization favors both a large final distinguishability and rapid signal accumulation throughout the readout window.
The offset $\epsilon>0$ ensures numerical stability of the logarithms when $\mathrm{SNR}(t)$ is small. 
Physical feasibility is enforced separately through penalty terms $P_j$ .

To suppress parasitic nonlinear effects, the photon number penalty, $P_{N}$, prevents the intracavity field from exceeding the critical photon number $N_{\mathrm{max}}$:
\begin{equation}
P_{N} = \frac{1}{t_f} \int_0^{t_f} \left[ \max\left(0, \frac{|\alpha(t)|^2}{N_{\mathrm{max}}} - 1\right) \right]^2 dt.
\end{equation}

A soft pulse-area constraint is further introduced as $P_{\mathrm{area}}$, to regularize the overall pulse area of the control envelope. This term constrains the integrated amplitude of $g_c(t)$ to remain close to that of the physical seed, preventing the agent from exploring physically unreasonable high-energy solutions:
\begin{equation}
P_{\mathrm{area}} = \left( \frac{\int_0^{t_f} |g_c(t)| \mathrm{d}t}{A_{\mathrm{seed}}} - 1 \right)^2,
\end{equation}
where $A_{\mathrm{seed}}$ denotes the integrated area of the STA seed waveform.

With these physics-motivated penalties and the STA warm start, PPO’s clipped updates provide a conservative refinement mechanism that stabilizes learning within the feasible region.


\section{Numerical Simulations and Results}\label{s3}

This section benchmarks the proposed RL framework for longitudinal readout pulse design against a fixed analytical baseline. As the reference protocol, we use the polynomial seed obtained from the STA construction in \cite{PhysRevApplied.18.034010}:
\begin{equation}
g_c^{(\mathrm{seed})}(t)= -\,\frac{70\pi\, g_{z0}}{\kappa\, t_f^{7}}\, t^{3}\,(t-t_f)^{3},
\label{eq:gc_seed_poly}
\end{equation}
which satisfies smooth turn-on/off boundary conditions and provides a reproducible comparison point. The PPO-optimized protocol is evaluated in terms of training convergence, pointer-state dynamics, SNR improvement, and robustness to parameter variations.

All simulations use the effective model introduced in Sec.~\ref{s2.1}. Unless otherwise stated, the resonator parameters are fixed to $\kappa/2\pi = 1~\mathrm{MHz}$ and $\omega_r/2\pi = 6.6~\mathrm{GHz}$, with a short readout window $t_f = 6\pi/(100\kappa)\approx 30~\mathrm{ns}$. 

\begin{figure*}[hbtp]
    \centering
    \includegraphics[width=1\textwidth]{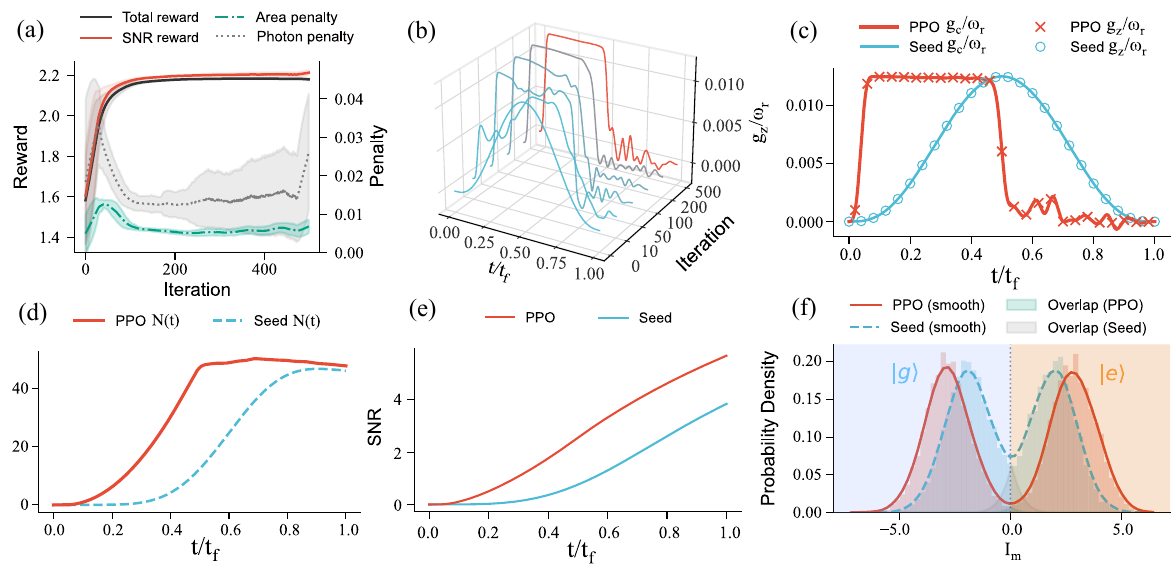}
    \caption{Physics-seeded PPO: training dynamics and readout performance.
        (a) Mean total reward $R(\mathbf{c})$ (black solid) and mean SNR reward $R_{\mathrm{SNR}}(\mathbf{c})$ (red solid) versus training iteration.
        The dashed and dotted lines show the mean penalty terms $P_{\mathrm{area}}$ and $P_{N}$, respectively.
        Shaded areas represent the standard deviation across independent training runs.
        (b) Colors indicate the optimization iteration of $g_c(t)$ (from early purple to late yellow), showing the evolution of the normalized control pulse $g_c(t)/\omega_r$ over the normalized time $t/t_f$.
        (c) Auxiliary envelope $g_c(t)$ (solid lines) used to drive the effective mean-field dynamics, and the corresponding physical coupling $g_z(t)$ (markers) reconstructed via Eq.~(\ref{eq:inverse_engineering}) in Sec.~\ref{s2.1}. Blue denotes the STA seed and red denotes the PPO result.
        (d) Intracavity photon number $N(t)=|\alpha(t)|^2$ for the STA seed (blue dashed) and the PPO protocol (red solid). The horizontal reference indicates the imposed photon-number limit $N_{\mathrm{max}}=50$.
        (e) Cumulative SNR of the normalized readout time $t/t_f$ for the STA seed (blue line) and the PPO-optimized pulse (red line).
        (f) Histograms of the integrated homodyne record $I_m$ for the qubit states $|g\rangle$ and $|e\rangle$. Blue dashed lines and red solid lines (with shaded fill) correspond to the distributions obtained using the STA seed and the PPO protocol, respectively.}
    \label{fig:2}
\end{figure*}

To stay within the validity regime of the linear model, the overall drive amplitude is normalized by enforcing a photon-number limit. Starting from a reference scale $g_{z0}^{(\mathrm{base})}/2\pi = 21~\mathrm{MHz}$~\cite{PhysRevApplied.18.034010}, the STA seed reaches a peak photon number $N_{\mathrm{seed}}\approx 1072$ under the linear dynamics. Imposing $N_{\mathrm{max}}=50$, the waveform is uniformly rescaled by
\begin{equation}
s=\sqrt{\frac{N_{\mathrm{max}}}{N_{\mathrm{seed}}}},\qquad
g_{z0}= s\, g_{z0}^{(\mathrm{base})}\approx 4.54~\mathrm{MHz},
\end{equation}
using the linear scalings $\alpha(t)= s\,\alpha^{(\mathrm{base})}(t)$ and $N(t) = s^2 N^{(\mathrm{base})}(t)$. This normalization places both the STA baseline and the RL search space on the same feasible scale, enabling a fair comparison under identical constraints.

For numerical integration, the interval $[0,t_f]$ is discretized with a uniform grid of $N_{\mathrm{grid}}=5001$ points, which is sufficient for converged cavity trajectories and reward metrics.

For numerical evaluation, we compute the SNR using the compact form
\begin{equation}
\mathrm{SNR}(t)
=
\frac{\sqrt{\eta\kappa}\int_{0}^{t} d(\tau)\,\mathrm{d}\tau}{\sqrt{S_{\mathrm{eff}}\,t}},
\label{eq:snr_compact}
\end{equation}
which is obtained from the quantum definition in Eq.~(\ref{eq:snr_quantum_def}) by using the input-output relation. 
Within the linear model, the conditional cavity states remain coherent so that the noise variance is displacement-independent. 
Details above are given in Appendix~\ref{app:snr_proof}.

A central feature of the proposed framework is that the policy search is guided by physically motivated initialization and explicit feasibility constraints. 
Figure~\ref{fig:2}  summarizes the training behavior of the PPO agent. 
Specifically, the physical start point described in Sec.~\ref{s2.3} initializes the policy around the STA seed coefficients, so that early exploration remains within a physically plausible region of the control landscape.

As shown in Fig.~\ref{fig:2}(a), both the reward and the resulting $\mathrm{SNR}(t_f)$ improve rapidly during the initial stage of training and then gradually saturate. 
The remaining fluctuations arise primarily from stochastic policy sampling and finite-batch estimation in PPO. 
Figure~\ref{fig:2}(b) provides a compact view of how the learned waveform evolves: starting from a smooth seed-like profile at early iterations, the agent progressively reshapes the control into a flat-top-like protocol that rapidly moves toward the constraint boundary and maintains a high-amplitude plateau over most of the readout window. 
These results indicate that enforcing feasibility at the parameterization level improves training stability and sample efficiency when optimizing readout performance under multiple rigid constraints.

To elucidate the physical mechanism behind the performance improvement achieved by PPO under rigid constraints, we compare the control waveforms and the resulting intracavity photon-number dynamics for the STA seed and the PPO-optimized solution.

Figure~\ref{fig:2}(c) shows that the STA seed, restricted by the polynomial ansatz, ramps up more gradually and reaches its effective operating amplitude later in the readout window. 
By contrast, PPO reshapes the envelope so that the drive reaches the allowed range rapidly at early times while remaining smooth due to the B-spline parameterization and boundary conditions. Due to its smoothness-preserving property, the corresponding physical pulse $g_z(t)$ closely tracks the $g_c(t)$.

This difference is directly reflected in the photon-number trajectories in Fig.~\ref{fig:2}(d). The PPO-optimized pulse drives $N(t)$ to approach the bound $N_{\mathrm{max}}$ quickly and keeps it close to the bound over most of the integration window, whereas the STA seed spends a longer fraction of the window at suboptimal photon number. 
Since the measurement signal accumulation in our effective model increases with the pointer separation (and hence with $|\alpha(t)|$), operating near the photon-number limit for a larger portion of the readout time yields a higher integrated SNR at fixed $t_f$. This constraint-saturating, hold-like behavior constitutes the main mechanism by which PPO outperforms the analytical STA baseline under the same rigid limits.

\begin{figure}[b]
    \centering
    \includegraphics[width=0.42\textwidth]{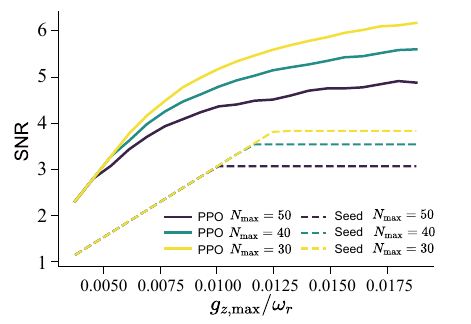} 
    \caption{Scalability of the final-time SNR under dual hardware constraints.
    Final $\mathrm{SNR}(t_f)$ as a function of the maximum allowable longitudinal coupling $g_{z,\max}/\omega_r$.
    Different colors correspond to photon-number caps $N_{\max}\in\{30,40,50\}$.
    Solid lines: PPO-optimized protocols. Dashed lines: analytical STA baseline.}
    \label{fig:scalability}
\end{figure}

We quantify the readout improvement using the time-resolved SNR defined in Eq.~(\ref{eq:snr_compact}). 
In Fig.~\ref{fig:2}(e), the PPO strategy achieves an absolute SNR enhancement of approximately $50\%$ compared to the baseline (increasing from $\approx 3.8$ to $\approx 5.7$). Notably, the PPO curve exhibits a much steeper slope in the early stage, indicating a faster information acquisition rate. This is a direct result of the ``Saturate-and-Hold'' photon dynamics observed in Fig.~\ref{fig:2}(d), which effectively maximizes the instantaneous signal magnitude over the full duration of the pulse.

To further validate the readout performance, Fig.~\ref{fig:2}(f) presents the simulated histograms of the integrated homodyne record $I_m$. The distributions for the ground state $|g\rangle$ and excited state $|e\rangle$ are well approximated by Gaussians in the linear coherent-state regime. The optimization effect is visually evident: the PPO protocol pushes the centers of the two distributions significantly further apart compared to the STA seed. 
Consequently, the overlap area between the two distributions which corresponds to the readout error probability is drastically reduced. 
This geometric separation confirms that the PPO-optimized pulse translates the theoretical SNR gain into a reduced single-shot assignment error.

In realistic quantum processors, achievable readout performance is often bounded by multiple hardware limits simultaneously. Figure~\ref{fig:scalability} benchmarks the scalability of the proposed framework under two representative constraints: an upper bound on the maximum physical longitudinal coupling amplitude $g_{z,\max}$ and a limit on the peak intracavity photon number $N_{\max}$.

\begin{figure}[b]
    \centering
    \includegraphics[width=0.45\textwidth]{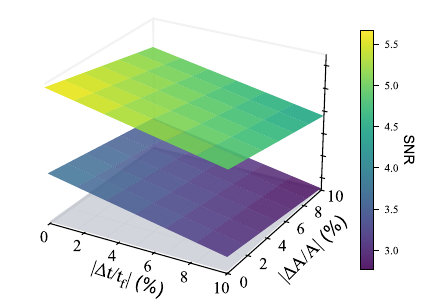} 
    \caption{Worst-case robustness to timing and amplitude errors. The upper surface corresponds to the PPO-optimized pulse and the lower surface to the STA seed. The final SNR is evaluated under bounded uncertainties in a timing error $|\Delta t|/t_f|$ (x-axis) and a multiplicative amplitude error $|\Delta A|/A|$ (y-axis). For each error bound, we report the worst-case SNR, defined as the minimum SNR over all error realizations within $\delta_t\in[-|\Delta t|,|\Delta t|]$ and $\delta_A\in[-|\Delta A|,|\Delta A|]$. }
    \label{fig:robustness}
\end{figure}

The comparison highlights a distinct advantage of the RL approach. For the STA baseline, increasing $g_{z,\max}$ beyond a moderate level yields diminishing returns: once the protocol already approaches the photon-number limit, further enlarging the coupling bound does not substantially increase the time-integrated measurement signal within the fixed readout window. 
In contrast, the PPO-optimized protocols adapt their waveform shape as the coupling budget increases. In particular, with a larger $g_{z,\max}$ the learned pulses can reach the photon-number limit earlier and maintain near-bound photon population over a larger fraction of the window, while still satisfying the endpoint and smoothness constraints. Consequently, the final $\mathrm{SNR}(t_f)$ continues to improve with $g_{z,\max}$ over a broader range of coupling limits, especially for tighter photon limits.

Fig.~\ref{fig:scalability} demonstrates that the proposed RL framework can empirically map out high-performing protocols near the effective performance boundary set by simultaneous amplitude and photon-number constraints, thereby making more efficient use of the available hardware resources than fixed-ansatz baselines.

The robustness trends in Fig.~\ref{fig:robustness} highlight a clear separation between the two design strategies. Across the explored uncertainty range, the PPO-optimized waveform exhibits a larger high-performance region and a slower degradation of the readout metric as the admissible errors increase. In contrast, the STA seed shows a noticeably steeper sensitivity to timing mismatch and amplitude miscalibration, indicating a tighter operating window. From an experimental standpoint, this improved tolerance translates into reduced calibration overhead and more reliable performance under routine day-to-day fluctuations.

\section{Discussion and conclusions}\label{s4}

The numerical results presented herein indicate that the PPO agent spontaneously converges to a ``flat-top'' readout pulse, yielding approximately $50\%$ improvement in the final SNR (increasing from 3.8 to 5.7) compared to the STA baseline. 
Geometrically, this strategy reshapes the cavity phase-space trajectory from the ``triangle-like'' profile typical of STA to a ``rectangle-like'' shape. 
By driving the intracavity photon number to approach $N_{\mathrm{max}}$ rapidly and sustaining it near the limit for the majority of the readout window, the agent maximizes the integrated signal area. Notably, this``saturate-and-hold'' protocol is automatically discovered by the agent under strict constraints, mirroring the empirical ``cavity ringing'' intuition in dispersive readout, yet interpreted here as an optimal compromise lying on the boundary of the feasible region spanned by $g_{\max}$ and $N_{\mathrm{max}}$.

Methodologically, the results validate the efficiency of the global B-spline parameterization. This choice is dictated by the inverse-engineering map, $g_z(t)=g_c(t)+\omega_r^{-2}\ddot{g}_c(t)$: discrete waveform generation would introduce high-frequency noise, which is catastrophically amplified by the second derivative term. Cubic B-splines provide a $C^2$-smooth parameterization that inherently enforces bandwidth limits. Furthermore, by reducing the control dimensionality from dense time grids ($N\sim5000$) to sparse spline coefficients ($\mathcal{O}(10)$), our approach significantly lowers the optimization complexity. This low-dimensional, physics-structured action space constrains exploration to feasible regions, thereby accelerating convergence and preventing pathological, high-energy solutions common in black-box optimization.

The robustness analysis reveals a critical advantage for experimental deployment. The PPO-optimized pulses sustain superior SNR levels across a broad range of parameter mismatches, exhibiting a robust high-performance plateau compared to the STA seed. This observation suggests a practical two-step calibration workflow: using the analytical STA pulse as a physics-based initial guess, followed by closed-loop RL fine-tuning. Such a parameterized search can compensate for device-specific distortions not captured by the linear model, while the enhanced robustness relaxes the stringent precision requirements typically needed for fast readout calibration.

In conclusion, while the proposed RL framework yields sizable quantitative gains, bridging to hardware deployment will require relaxing idealizations such as the single-mode approximation and developing robust transfer protocols against measurement noise. Nevertheless, the results support a physics-informed RL paradigm: by anchoring learning to a low-dimensional, smooth, and analytically initialized parameter space, we can effectively solve constrained quantum-control problems where traditional high-dimensional optimization is inefficient.

\section*{Acknowledgments}
Y.-H.C. was supported by the National Natural Science Foundation of China under Grant No. 12304390 and 12574386, the Fujian 100 Talents Program, and the Fujian Minjiang Scholar Program. Y. X. was supported by the National Natural Science Foundation of China under Grant No. 62471143, the Key Program of National Natural Science Foundation of Fujian Province under Grant No. 2024J02008.

\section*{Data availability}
The data that support the findings of this study are available from the corresponding author upon reasonable request.

\section*{Code availability}
Custom code used in this study is available from the corresponding author upon reasonable request.

\section*{Author contributions}
Y.-H.C. and Y.X. conceived the idea and supervised the project. Y.Y. performed the theoretical derivations and numerical simulations and wrote the manuscript. Y.Q. and X.Z. assisted with the data analysis and checked the validity of the results. All authors discussed the results and contributed to the final version of the manuscript.

\appendix

\section{Derivation of the Inverse Engineering Control Law}
\label{app:inverse_engineering}

This appendix derives the inverse-engineering relation used in the main text,
Eq.~(\ref{eq:inverse_engineering}).

Starting from the longitudinal Hamiltonian in Eq.~(\ref{eq:hamiltonian}), we use the resonator-frame Hamiltonian (equivalently, the interaction picture with respect to $\omega_r \hat{a}^\dagger \hat{a}$, which leaves the coupling term unchanged).
\begin{equation}
    H(t) = \omega_r \hat{a}^\dagger \hat{a} + g_z(t)\sigma_z (\hat{a}^\dagger + \hat{a}).
\end{equation}
To accelerate the QND measurement process and eliminate the non-adiabatic transitions caused by the time-dependent coupling $g_z(t)$, we seek a solution to the time-dependent Schrödinger equation $i\partial_t |\Psi(t)\rangle = H(t)|\Psi(t)\rangle$ in the form of a parameterized ansatz:
\begin{equation}
    |\Psi(t)\rangle = e^{-i\Theta(t)} \mathcal{V}(t)|\phi(t)\rangle,
\end{equation}
where $|\phi(t)\rangle$ represents the eigenstate of the unperturbed harmonic oscillator (free evolution), and $\Theta(t)$ is a global phase factor. The key component is the time-dependent unitary transformation $\mathcal{V}(t)$, which is constructed to disentangle the longitudinal coupling. Based on the STA formalism, we adopt the displacement-operator ansatz:
\begin{equation}
    \mathcal{V}(t) = \exp\left[ -i \frac{\dot{g}_c(t)}{\omega_r^2} \sigma_z (\hat{a}^\dagger + \hat{a}) \right] \exp\left[ -\frac{g_c(t)}{\omega_r} \sigma_z (\hat{a}^\dagger - \hat{a}) \right].
\end{equation}
Here, the two exponential terms correspond to momentum and position displacement operations in phase space, parameterized by an auxiliary function $g_c(t)$ and its derivative $\dot{g}_c(t)$.

Substituting this ansatz into the Schrödinger equation, the condition for $|\Psi(t)\rangle$ to be an exact solution is that the parameters satisfy the classical equation of motion for a forced oscillator. Specifically, by demanding that the effective Hamiltonian governing $|\phi(t)\rangle$ reduces to the free oscillator form, we obtain the following dynamical constraint:
\begin{equation}
    \ddot{g}_c(t) + \omega_r^2 [g_c(t) - g_z(t)] = 0.
\end{equation}
This second-order differential equation connects the auxiliary trajectory $g_c(t)$ to the physical control pulse $g_z(t)$. Rearranging this equation yields the inverse engineering control law:
\begin{equation}
    g_z(t) = g_c(t) + \frac{1}{\omega_r^2}\ddot{g}_c(t).
\end{equation}
This result demonstrates that by designing a smooth auxiliary trajectory $g_c(t)$ (satisfying $g_c=\dot g_c=\ddot g_c=0$ at $t=0$ and $t=t_f$), one can analytically reconstruct the required physical coupling strength $g_z(t)$ to achieve high-fidelity readout.

\section{Equivalence of the semi-classical and quantum SNR definitions}
\label{app:snr_proof}

In this appendix, we justify the semi-classical SNR expression used in the main text,
Eq.~(\ref{eq:snr_compact}), by connecting it to the standard quantum continuous-measurement
definition based on homodyne detection. 
The key point is that, within the linearized
effective model adopted here, the resonator field remains (approximately) in a coherent
state throughout the readout.

\subsection{Quantum homodyne observable and integrated measurement record}

We consider homodyne detection of the output field quadrature. The instantaneous
homodyne observable can be written as
\begin{equation}
\hat{I}_\phi(t) = \sqrt{\eta}\,
\Bigl[e^{-i\phi}\hat{a}_{\mathrm{out}}(t)+e^{i\phi}\hat{a}_{\mathrm{out}}^\dagger(t)\Bigr],
\label{eq:homodyne_current_operator}
\end{equation}
where $\eta$ is the overall detection efficiency and $\phi$ is the local-oscillator phase.
The output field satisfies the input--output relation
\begin{equation}
\hat{a}_{\mathrm{out}}(t)=\sqrt{\kappa}\,\hat{a}(t)-\hat{a}_{\mathrm{in}}(t),
\label{eq:io_relation}
\end{equation}
with $\hat{a}(t)$ the intracavity field operator and $\hat{a}_{\mathrm{in}}(t)$ the input field
(vacuum in our setting).

The boxcar-integrated measurement operator is then
\begin{equation}
\hat{\mathcal{M}}_\phi(t) \equiv \int_{0}^{t}\hat{I}_\phi(\tau)\,d\tau.
\label{eq:int_measurement_operator}
\end{equation}
For distinguishing two qubit states $s\in\{e,g\}$, the quantum SNR is naturally defined as
\begin{equation}
\mathrm{SNR}_{\mathrm{Q}}(t)=
\frac{\left|\langle \hat{\mathcal{M}}_\phi(t)\rangle_e-\langle \hat{\mathcal{M}}_\phi(t)\rangle_g\right|}
{\sqrt{\mathrm{Var}_e[\hat{\mathcal{M}}_\phi(t)]+\mathrm{Var}_g[\hat{\mathcal{M}}_\phi(t)]}},
\label{eq:snr_quantum_def1}
\end{equation}
where $\mathrm{Var}_s[\hat{\mathcal{M}}]\equiv \langle \hat{\mathcal{M}}^2\rangle_s-\langle \hat{\mathcal{M}}\rangle_s^2$.

\subsection{Signal separation (numerator)}

In the longitudinal readout considered here, the qubit state conditions the cavity
displacement, and within the linear effective model the cavity remains (approximately)
coherent for each qubit state:
\begin{equation}
\alpha_s(t)\equiv \langle \hat{a}(t)\rangle_s,\qquad s\in\{e,g\}.
\end{equation}
For vacuum input, $\langle \hat{a}_{\mathrm{in}}(t)\rangle = 0$, so Eq.~(\ref{eq:io_relation}) gives
\begin{equation}
\langle \hat{a}_{\mathrm{out}}(t)\rangle_s=\sqrt{\kappa}\,\alpha_s(t).
\end{equation}
Choosing $\phi$ to align the measured quadrature with the pointer-state displacement axis,
the instantaneous signal contrast becomes proportional to the pointer separation
\begin{equation}
d(t)\equiv |\alpha_e(t)-\alpha_g(t)|.
\end{equation}
Accordingly, the integrated signal separation reads
\begin{equation}
\left|\langle \hat{\mathcal{M}}_\phi(t)\rangle_e-\langle \hat{\mathcal{M}}_\phi(t)\rangle_g\right|
= \sqrt{\eta\kappa}\int_0^t d(\tau)\,d\tau,
\label{eq:signal_separation}
\end{equation}
which matches the numerator used in Eq.~(\ref{eq:snr_quantum_def}) in the main text.

\subsection{Noise scaling and the effective noise spectral density $S_{\mathrm{eff}}$ (denominator)}

The noise in homodyne detection originates from vacuum fluctuations entering through
$\hat{a}_{\mathrm{in}}(t)$ (and, more generally, from effective vacuum due to inefficiency and
added technical noise). A central property of coherent states is that their quadrature
fluctuations are identical to those of vacuum and do not depend on the displacement
$\alpha_s(t)$.
\begin{equation}
\mathrm{Var}_e[\hat{\mathcal{M}}_\phi(t)]\simeq \mathrm{Var}_g[\hat{\mathcal{M}}_\phi(t)].
\end{equation}

Specifically, we model the measurement record as
\begin{equation}
I(t)=\sqrt{\eta\kappa}\,d(t)+\xi(t),
\end{equation}
where $\xi(t)$ is a zero-mean Gaussian white noise process with
\begin{equation}
\langle \xi(t)\xi(t')\rangle = S_{\mathrm{eff}}\delta(t-t').
\label{eq:Seff_def_appendix}
\end{equation}
Here $S_{\mathrm{eff}}$ absorbs the quadrature normalization and vacuum-noise prefactors
(as well as any additional effective noise beyond ideal vacuum), so that the boxcar-integrated
noise obeys the compact relation
\begin{equation}
\mathrm{Var}\!\left[\int_0^t \xi(\tau)\,d\tau\right]=S_{\mathrm{eff}}\,t.
\label{eq:noise_variance_appendix}
\end{equation}

\subsection{Result: recovery of the semi-classical SNR expression}

Combining the signal separation in Eq.~(\ref{eq:signal_separation}) with the noise scaling
in Eq.~(\ref{eq:noise_variance_appendix}), the resulting SNR takes the form
\begin{equation}
\mathrm{SNR}(t)=
\frac{\sqrt{\eta\kappa}\int_0^t d(\tau)\,d\tau}{\sqrt{S_{\mathrm{eff}}\,t}},
\end{equation}
which is exactly Eq.~(\ref{eq:snr_compact}) in the main text.

Thus, within the coherent-state and effective-vacuum-noise limit of the linear model, maximizing Eq.~(\ref{eq:snr_compact}) is equivalent to maximizing the standard homodyne distinguishability.

\section{Sample-efficiency benchmark via iterations-to-target SNR}
\label{app:2}
\renewcommand{\thefigure}{S\arabic{figure}}
\setcounter{figure}{0}

\begin{figure}[b]
    \centering
    \includegraphics[width=0.4\textwidth]{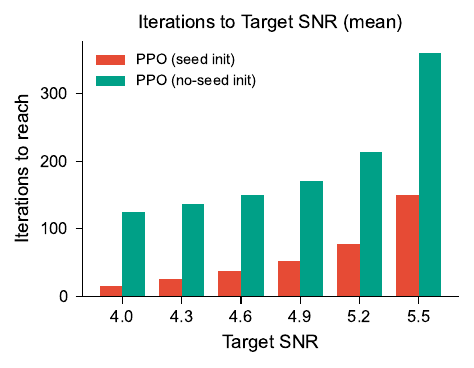} 
    \caption{Mean PPO iterations required to reach a target SNR. Bars compare seeded and no-seed initialization across several target SNR thresholds. Seeded initialization consistently reduces the iterations-to-target, demonstrating improved sample efficiency.}
    \label{fig:s1}
\end{figure}

To quantify the sample efficiency of PPO under different initializations, we measure the number of training iterations required to reach a prescribed target SNR. Figure~\ref{fig:s1} reports the mean iterations-to-target for a set of target SNR thresholds, comparing seeded initialization (policy initialized near a physics-motivated baseline) and no-seed initialization (generic initialization).

For each independent run, we define the iterations-to-target as the first PPO iteration at which the evaluation SNR reaches (or exceeds) the target threshold; we then average this quantity over multiple random seeds. Across all targets shown, seeded initialization consistently requires fewer iterations, demonstrating substantially improved training efficiency. Across all target thresholds, seeded initialization consistently reduces the iterations-to-target, indicating that the physics-informed seed provides a systematically advantageous starting point for PPO. 

This bias guides exploration toward high-performing regions of the control landscape, enabling the agent to reach a given SNR faster and more reliably under the same training budget, with the advantage persisting even for the highest targets considered.

	\bibliography{reference}
\end{document}